\documentclass[conference]{IEEEtran}

\usepackage{graphicx}
\usepackage{epstopdf}
\usepackage{url}
\usepackage{booktabs}
\usepackage{xspace}
\usepackage{tabularx}
\usepackage{microtype}

\newcommand{\sname}{TrustBase}

\newcommand{\name}{TrustBase\xspace}

\begin{document}

\title{\name : An Architecture to Repair and Strengthen Certificate-based Authentication}

\author{
{\rm Mark O'Neill\qquad Scott Heidbrink\qquad Scott Ruoti\qquad Jordan Whitehead}\\ {\rm Dan Bunker\qquad Kent Seamons\qquad Daniel Zappala}\\
Brigham Young University\\
{\small mto@byu.edu, sheidbri@byu.edu, ruoti@isrl.byu.edu, jaw@byu.edu,} \\{\small dbunked@gmail.com, seamons@cs.byu.edu, zappala@cs.byu.edu}
} 


\maketitle

\begin{abstract}
We describe \name, an architecture that provides certificate-based authentication
as an operating system service. \name enforces best practices for certificate validation for all applications
and transparently enables existing applications to be strengthened against failures of the CA system.
The \name system allows simple deployment of authentication systems that harden the CA system.
This enables system administrators, for example, to require certificate revocation checks on all TLS connections,
or require STARTTLS for email servers that support it.
\name is the first system that is able to secure all TLS traffic, using an approach compatible with all operating systems.
We design and evaluate a prototype implementation of \name on Linux, evaluate its security,
and demonstrate that it has negligible overhead and universal compatibility with applications.
To demonstrate the utility of \name, we have developed six authentication services that strengthen certificate validation
for all applications.
\end{abstract}

\section{Introduction}

Server authentication on the Internet currently relies on the certificate authority (CA) system. 
While establishing a secure SSL/TLS session with a server, clients receive a digital certificate that vouches for the identity of the server.
This certificate is signed by a certificate authority, and the signature is validated against a list of certificate authorities that is shipped with the client or operating system. 
This provides assurance that the client is connected to a legitimate server and not one controlled by an attacker.

Unfortunately, certificate validation is challenged by three significant problems:
\begin{itemize}

\item \textbf{Applications frequently do not properly validate the server's certificate~\cite{georgiev2012most,fahl2012eve,brubaker2014using}.}
  This is caused by failure to use validation functions, incorrect usage of libraries, and also developers who disable validation during development and forget to enable
  it upon release~\cite{fahl2013rethinking}. These implementation mistakes deceive the user into believing
  her session is secure, while leaving the application vulnerable to man-in-the-middle (MITM) attacks.

\item \textbf{The CA system is vulnerable to being hijacked even when applications are implemented correctly.}
  This is largely due to the fact that most CAs are able to sign certificates for any host, reducing the strength of the CA system
  to that of the weakest CA~\cite{eckersley2010observatory}.
  This weakness was exploited in 2011 when DigiNotar's servers were hacked and more than 500 certificates were
  fabricated by the intruder, including a certificate for Gmail that allowed the intruder to access stored email for 300,000 Iranians \cite{iran}.
  This happened despite the fact that Gmail does not use DigiNotar to sign its certificates.
  This problem is exacerbated by CAs that do not follow best practices~\cite{marlinspike2011ssl,durumeric2013analysis} and
  governmental ownership and access to CAs~\cite{eckersley2011decentralized, soghoian2012certified}.
  
\item \textbf{Improvements to the CA system have difficulty being widely deployed.}
  Protocols to enhance the certificate validation process have been developed, but the majority of applications have not yet integrated these improvements.
  Even relatively simple fixes, such as certificate revocation, are beset with problems \cite{revocation}.
  There is no widely-adopted platform assisting the deployment of improvements to the CA system,
  meaning researchers and developers have to individually modify applications to make advances, which
  severely limits their deployment.

\end{itemize}

In this paper, we introduce \name, an architecture for certificate authentication that solves each of these three problems.
\name is designed to (1) secure existing applications, (2) strengthen the CA system, and (3) provide simple 
deployment of improved authentication systems.

To address these problems, \name complements the existing certificate validation performed by applications
with additional validation methods that are enforced by the operating system. This approach
has several advantages. First, it protects against insecure applications---too much evidence shows that
developers make mistakes, and this will only get more
difficult as additional authentication methods such as Certificate Transparency \cite{rfc6962,ryan2014enhanced},
pinning \cite{evans2011certificate,tack}, and DANE \cite{hoffman2011using} become more widely available.
Second, it transparently enables existing applications to be strengthened against failures of the CA system.
For example, a browser that validates the EV cert of a bank is doing the best it currently can,
but it is still vulnerable to a CA that is hacked, allowing a man-in-the middle (MITM) to present fake but valid certificates.
\name provides complementary authentication services that
protect against these situations, such as checking notaries to ensure other hosts across the Internet
are seeing the same certificate for the bank.\footnote{Certificates for a notary service can be pinned in advance, so they
are not vulnerable to the MITM.}
Third, it enables a system administrator to 
ensure best security practices are properly followed.
\name allows the administrator for an organization to enforce consistent certificate validation across all applications,
such as requiring revocation checking or mandating that pinning is used to protect against MITM attacks.
\name provides administrators with a choice of authentication services that can
be used to harden the CA system.\footnote{We expect \name would be distributed with default settings by operating system vendors so that ordinary users would
  not need to configure it, but knowledgeable system administrators would be able to exercise greater control.}

We identify a number of design goals for \name, including full application coverage, universal deployment, and negligible overhead.
To meet these goals, \name uses traffic interception between the socket layer and the transport layer, an approach that is general enough to work on all major operating systems, both desktop and mobile. \name detects the initiation of TLS connections, finds the handshake
information, validates the server's certificate using a variety of configurable authentication systems, and then allows or blocks the
connection based on the results of this additional validation.
This allows \name to harden the certificate validation of
all applications, regardless of which libraries they use. For new or
modified applications, \name provides a simple certificate validation API that can be called directly. 

Our contributions include:

\begin{itemize}

\item {\bf An architecture for enforcing best practices for 
  certificate validation on all applications:} \name requires standard certificate
  validation procedures and optionally adds additional authentication services, both of which are
  enforced by the operating system and controlled
  by the administrator. This repairs broken validation for poorly-written applications
  and optionally strengthens the validation done by well-written applications.
  Applications are allowed to be more strict, but not less strict than \name in their validation decisions.

\item {\bf Simplified deployment of authentication systems that strengthen the CA system:} \name provides a
  plugin API and a policy engine that dynamically loads authentication
  systems based on administrator preferences. They
  can be written in either C or Python, with the ability to add
  support for additional languages. The administrator can use policies
  to define how multiple authentication systems cooperate, for example
  using unanimous consent or threshold voting.

\item {\bf A research prototype of \name for Linux:} We develop
  a loadable kernel module that provides general traffic interception
  and TLS handling for Linux. This module communicates via the Netlink API to
  the policy engine residing in user space for parsing and validation of certificates.
  We describe how this same architecture can be implemented
  on Windows, Mac OS X, iOS, and Android devices.

\item {\bf A security analysis of \name:} We provide a security analysis of \name, including its centralization, application coverage, and
  the hardening we have done on the Linux implementation. We provide a threat analysis and demonstrate how \name can
  thwart attacks that include a hacked CA, a subverted local root store, and a STARTTLS downgrade attack. We also demonstrate the ability
  of \name to fix applications that do not validate hostnames or skip certificate validation altogether.

\item {\bf An evaluation of \name:} We evaluate the \name prototype for
  performance, compatibility, and utility. (1) We show that \name has
  negligible performance overhead, with no measurable impact on
  latency. (2) We demonstrate that \name enforces correct certificate
  validation on all popular Linux libraries and tools and on the most
  popular Android applications. (3) We describe
  six authentication services that we have developed and report on
  how simple and straightforward it was to develop these services in \name.

\end{itemize}

\section{Related Work}

The typical goal of an attacker who wishes to subvert TLS is to perform a MITM attack on unsuspecting clients,
substituting his own certificate in place of the original. In this section we discuss flawed implementations, weaknesses of the CA system that make these attacks possible, and various alternatives to the CA system that have been developed.
We then discuss related projects that try to address the problems affecting applications and the CA system.

\subsection{Flawed Implementations}

Several studies demonstrate widespread vulnerabilities in client-side validation of certificates, in both mobile and desktop environments.
Georgiev et al.~\cite{georgiev2012most} discovered that many critical software applications outside the  browser rely on completely broken TLS validation libraries, primarily due to poorly designed APIs.
Brubaker et al.~\cite{brubaker2014using} conducted a large-scale experiment of TLS implementations using millions of ``frankencerts'', random certificates generated from portions of real certificates. 
They discovered hundreds of flaws in popular TLS libraries and browsers.
Fahl et al.~\cite{fahl2012eve} analyzed 13,500 popular Android apps  and determined that 8\% were vulnerable to TLS MITM attacks.
Lucky et al.~\cite{onwuzurike2015danger} experimented with 100 popular Android apps and found that several accept all certificates
and all hostnames, along with other unsafe practices.

There is also evidence that servers are often incorrectly setting up the certificate they send as part of the TLS handshake.
For example, Holz et al.~\cite{holz2011ssl} conducted a comprehensive study and determined that 
invalid certificate chains, certificate subjects, and self-signed certificates are
the source of many concerns.
Additionally, Vratonjic et al. revealed that a minority of the one million most popular web sites use certificates properly~\cite{vratonjic2013inconvenient}.


\subsection{Problems with the CA System}
Clark and van Oorschot~\cite{clark2013sok} provide an extensive survey of work examining the CA system and found that the CA system is surprisingly brittle.
If these problems with the CA system are not addressed, they will continue to allow malicious individuals, governments, and others to intercept, decrypt, and even modify TLS traffic.
This could have drastic consequences for user privacy, confidence in the safety of e-commerce, and national security.

Perhaps most troubling is that CAs are able to sign certificates for any server.
Mechanisms to restrict the signing power of certificate authorities are, in practice, unused.
For example, Durumeric et al.~\cite{durumeric2013analysis} discovered that there are over 1,800 certificate authorities that can issue certificates for any web site. CAs mismanage their private key and have been hacked multiple times, leaving forged certificates in the hands of the attackers \cite{marlinspike2011ssl}.
Moreover, some governments, including China and Russia, have their own CAs that
are trusted by all major browsers by default.
There has been some recent evidence that government institutions are coercing CAs into providing certificates for use in surveillance \cite{eckersley2011decentralized,soghoian2012certified}.
Finally, some companies have been found to be using their position as hardware vendors as a vector for secretly adding their own certificates to devices and then performing TLS MITM on those devices \cite{nokia,lenovo}.

\subsection{Authentication Improvements}

Due to these problems, there are a number of recent proposals to
improve or replace the current CA trust model. (1) Multi-path probing
such as Convergence \cite{marlinspike2011ssl} allows clients to
determine whether a server certificate they have received is different
from those seen by most other clients
\cite{wendlandt2008perspectives,alicherry2009doublecheck,holz2012x}.
Related systems use existing Certificate Authorities or centralized
notaries to vouch for the authenticity of a certificate
\cite{mecai,amann2012extracting, amann2012revisiting}.  (2) DNS-Based
Authentication of Named Entities (DANE)~\cite{hoffman2011using}
enables administrators to bind hostnames to their certificates,
permitting public keys to be transmitted via DNSSEC without involving
a CA.  (3) Certificate pinning~\cite{evans2011certificate,tack} allows
a web server to limit all future HTTPS connections to a limited set of
server certificates. (4) An audit log, such as Certificate
Transparency~\cite{rfc6962,ryan2014enhanced} or
the EFF Sovereign Keys project~\cite{sovereignkeys}, provides an
additional path to validate server certificates. The Accountable Key
Infrastructure (AKI)~\cite{kim2013accountable} extends this notion to
design a new infrastructure to validate public keys and reduce the
reliance on CAs.


\subsection{Related Projects}


Bates et al. created CertShim to address problems with TLS authentication~\cite{bates2014securing}.
Similar to \name, CertShim is an attempt to immediately fix TLS problems in existing apps and also support new authentication systems.
CertShim works by utilizing the \texttt{LD\_PRELOAD} environment variable to replace functions in dynamically-loaded security libraries with their own wrappers for those functions.
CertShim currently works in Linux and supports applications that use OpenSSL, libNSS, and older versions of GnuTLS and PolarSSL (with partial support for Java SSE).
Because CertShim relies on the \texttt{LD\_PRELOAD} variable, it doesn't provide the level of administrator control or full application
coverage we prefer, and has significant code maintenance issues. We discuss the detailed differences between CertShim and \name in Section~\ref{sec:discussion}.



Conti et al. created MITHYS to protect Android applications from MITM attacks~\cite{conti2013mithys}.
MITHYS employs a two stage approach: 
first, it attempts to MITM applications that are establishing TLS connections.
If the MITM is unsuccessful, MITHYS whitelists the app as secure and it will not attempt to MITM this app in the future.
If the MITM is successful, MITHYS will MITM all future TLS connections by the application.
Next, the MITM service checks whatever certificate chain it receives for the real TLS connection and checks it against a notary service hosted in the cloud.
If the chains do not match, then MITHYS will stop the connection, thereby preventing malicious MITM attacks.
MITHYS only works for HTTPS traffic, adds 
significant delays to all TLS connections that it protects (one to ten seconds), and 
only supports the current CA system. 
While additional authentication services could be added, it would have a significant impact on performance because it would be necessary to MITM all TLS connections.


Fahl et al. proposed a new framework for Android applications that would help developers correctly use TLS~\cite{fahl2013rethinking}.
Their work follows a similar principle as ours---instead of letting developers implement their own TLS code, TLS is a service provided by the operating system.
Application configuration files control TLS properties, including which authentication systems are used and any parameters to those systems (e.g., which attributes to pin).
Like \name, their work uses a pluggable framework for authentication schemes.
Fahl's approach is well-suited to mobile operating systems such as Android, where all applications are written in Java, but it is
difficult to extend this approach to operating systems that provide more general programming language support.
The \texttt{libtlssep} library on Linux, provides similar capabilities to Fahl's work and also isolates authentication
and session keys, making it more difficult for them to be exposed due to a program bug \cite{amour2015improving}. 
This approach provides good privilege separation, but it does not offer protection for existing applications or force adherence to administrator preferences.


\section{\name}

\name is motivated by the need to (1) secure existing applications, (2) strengthen the CA system, and (3) provide an easy path
to deploy alternative authentication systems.
In this section, we discuss the threat model, design goals, and architecture of the system.

\subsection{Threat Model}

Our threat model includes an active attacker that impersonates a remote host by providing a 
fake certificate during TLS server authentication. 
This includes remote hosts as well as MITM attackers located anywhere along the path to a remote host. It also includes local malware impersonating a remote host or attempting to subvert proper certificate validation.
The goal of the attacker is to establish a secure connection with the client. 

The application under attack may accept the fake certificate for the following reasons:
\begin{itemize}
\item The application has incorrect certificate validation procedures (e.g. limited or no validation) and the attacker exploits his knowledge of this to trick the application into accepting his fake certificate.
\item The attacker or malware managed to place a rogue certificate authority into the user's root store (or another trust store used by the application) so that he has become a trusted certificate authority. The fake certificate authority's private key was then used to generate the fake certificate used in the attack.
\item Non-privileged malware has altered or hooked security libraries the application uses to force acceptance of fake certificates (e.g. via malicious OpenSSL hooks and LD\_PRELOAD).
\item A legitimate Certificate Authority was compromised or coerced into issuing the fake certificate to the attacker.
\end{itemize}

Local attackers (malware) with root privilege are outside the scope of our threat model.
Attacks against the TLS protocol are also out of scope.

%

\subsection{Design Goals}


The design goals for \name are:

\begin{enumerate}

\item \textbf{Secure existing applications.}
  \name should override incorrect or absent certificate validation in current applications.

\item \textbf{Strengthen the CA system.} \name should permit using multiple
  authentication services to strengthen the validation provided by
  the CA system.

\item \label{goal:new} \textbf{Provide simple deployment of
  authentication systems.}  \name should facilitate the creation and
  deployment of new authentication systems.  Authentication systems
  should be easily developed and installed without requiring
  recompilation of \name.

\item \textbf{Full application coverage.}
  All certificates should be validated by {\sname}, including those provided to both existing applications and future applications.

\item \textbf{Universal deployment.}
  The {\sname} architecture should be designed to work on any major operating system including both desktop and mobile platforms.

\item \textbf{Negligible overhead.}
\name should have negligible performance overhead.
This includes ensuring that the user experience for applications is not affected, and also that applications do not time out because of the added overhead.

\end{enumerate}

\subsection{Architecture}

\begin{figure}
\centering
\includegraphics[width=0.8\columnwidth]{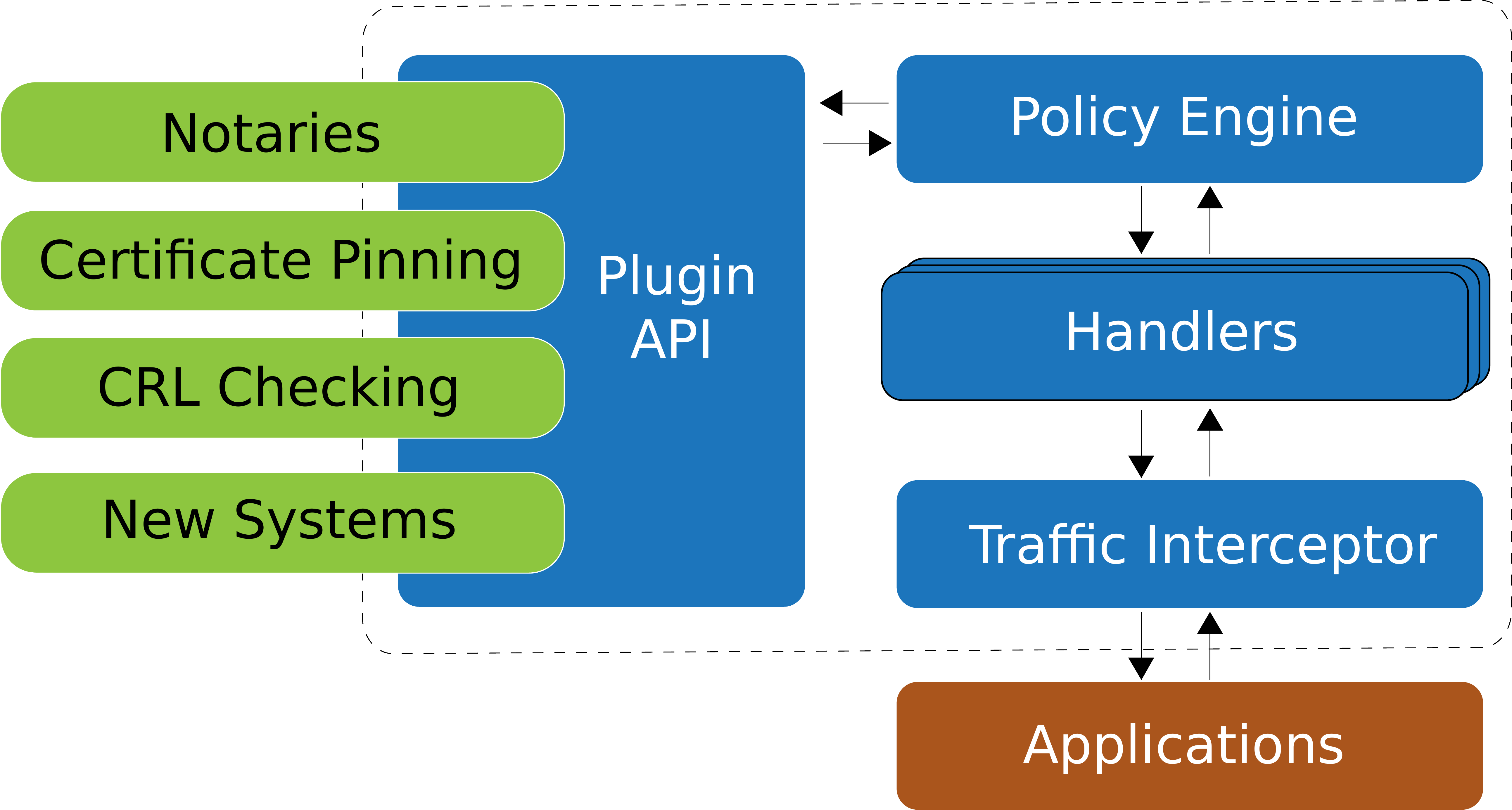}
\caption{\name architecture overview}
\label{architecture}
\end{figure}

The architecture for \name is given in Figure~\ref{architecture}.
The four principal components of \name are:

\begin{itemize}

\item
The \textbf{traffic interceptor} intercepts all network traffic and
delivers it to registered handlers for further processing. The
interceptor is generic, lightweight, and can provide traffic to any type of handler.

\item The \textbf{handlers} are state machines that examine a traffic stream to isolate data used for authenticating connections
  and then pass this data to the policy engine. Data provided to the policy engine includes everything from the relevent
  protocol that is intercepted.\footnote{This enables plugins to provide authentication methods that utilize TLS hello extensions, such as TACK \cite{tack}.} For example, with TLS this includes the \texttt{ClientHello} and \texttt{ServerHello} data, including the server certificate chain and the server hostname. The handler will allow or abort the connection, based on the policy engine's response.

\item
The \textbf{policy engine} is responsible for using the registered authentication system plugins to validate the server certificate extracted by the handler.
The policy engine also aggregates validation responses if there are multiple active authentication plugins.
The results of validation are sent back to the handler.
The policy is configured by the system administrator, with sensible operating system defaults for ordinary users.

\item
The \textbf{plugin API} defines functions that must be implemented by new authentication systems wishing to operate as plugins.
These plugins are provided with a server certificate chain and other contextual data and asked to validate the chain.
\name can dynamically load new authentication plugins without alteration or recompilation.

\end{itemize}

In addition to these main components, another module delivers operating system notifications that indicate when an application has been protected from an invalid certificate.
An API for the policy engine allows applications to be modified to invoke \name directly (see Section~\ref{sec:overriding}). 



\subsection{Traffic Interceptor}
\label{sec:design-traffic-interceptor}
The traffic interceptor monitors all network connections made by applications running on the system.
The interceptor is designed with the flexibility in mind, allowing for different handlers to be registered that can view and modify the intercepted traffic.
This flexibility can also be used by researchers who would like to leverage socket-level traffic interception for purposes outside of TLS validation; for example, measuring network traffic.
Traffic for any specific stream is intercepted only as long as a handler is interested in it. Otherwise, traffic is routed normally.

The traffic interceptor is needed to provide immediate and universal coverage for existing applications. If
a developer is willing to modify her application to call \name directly for certificate validation, then this can be avoided.

\subsection{Handlers}
\label{sec:secure-handshake-handler}

Handlers are state machines that examine a traffic stream to isolate data used for authenticating connections.
\name currently has both a TLS and an opportunistic encryption handler (e.g., STARTTLS), and due to the design of the traffic interceptor it is easy
to add support for new secure transport protocols as they become popular (e.g., QUIC, DTLS).

The TLS handler identifies a TLS handshake and extracts the \texttt{ClientHello} and \texttt{ServerHello} data, delivering
all of this data to the the policy engine, including the server's hostname and certificate chain. Upon notification from
the policy engine, the TLS handler will abort any TLS connections that do not have a valid certificate chain.

The TLS handler is a very simple state machine that functions according to the following flow:

\begin{enumerate}

\item Detect the creation of a new TCP connection.

\item Determine whether the connection is attempting to establish a TLS session. This detection is nearly always completed based on the first received payload, as TLS records have an easily recognizable header. If the connection is not establishing a TLS session, stop monitoring this connection.	

\item Allow the client to send its \texttt{ClientHello}, and then wait for the server to respond with its \texttt{ServerHello} and \texttt{Certificate} messages. While waiting for the server's certificate, cache any packets the server sends, and only forward these to the application after the connection is authenticated.

\item Send the server's host data (hostname, port, and IP address) and captured TLS handshake message data (including certificate chain) to the policy engine for validation.

\item Based on the policy engine's response, take appropriate action:\\

  \texttt{POLICY\_RESPONSE\_INVALID}: Receive a copy of the certificate from the policy engine which has its public key and signature scrambled, and then forward it to the application. This will prevent the application from establishing a TLS session even if the application blindly accepts the certificate and allows more verbose applications to display a relevant error message. \name also ensures that the connection is terminated. When applications are modified to call \name directly,
  they can receive more detailed error codes to display to the user. Operating system notifications can be used to inform users when applications are blocked by an invalid certificate.\\
	
\texttt{POLICY\_RESPONSE\_VALID}: Allow the connection to continue and stop monitoring it.
	

\end{enumerate}

The opportunistic encryption handler is also a simple state machine, which identifies when TLS begins on an plainttext connection and then hands the connection
to the TLS handler for additional processing. Both handlers are discussed in more detail in Section~\ref{linux-implementation}.

\subsection{Policy Engine}
\label{sec:design-policy-engine}

When the policy engine receives a validation request from the TLS handler, it will query each of the registered authentication system plugins to validate the server's certificate chain and host data.
Authentication system plugins can respond to this query in one of four ways:

\begin{itemize}
\item \texttt{PLUGIN\_RESPONSE\_VALID}: The plugin has determined that the certificate is valid for the given host.

\item \texttt{PLUGIN\_RESPONSE\_INVALID}: The plugin has determined that the certificate is {\em not} valid for the given host.

\item \texttt{PLUGIN\_RESPONSE\_ABSTAIN}: The plugin is unable to determine if the certificate is valid.

\item \texttt{PLUGIN\_RESPONSE\_ERROR}: The plugin encountered an error. This is also used if the query to the plugin times out.
\end{itemize}

Abstain and error responses are mapped to the valid or invalid responses, as
defined in a configuration file.

\name classifies plugins as either ``necessary'' or ``voting'', as defined in the configuration file. All plugins in the ``necessary'' category must indicate the certificate is valid, otherwise the policy engine will mark the certificate as invalid. If the necessary plugins validate a certificate, the responses from the remaining ``voting'' plugins are tallied. If the aggregation of valid votes is above a preconfigured threshold, the certificate is deemed valid by the policy engine.







The policy engine returns \texttt{POLICY\_RESPONSE\_VALID} or \texttt{POLICY\_RESPONSE\_INVALID} as appropriate.


A write-protected configuration file lists the plugins to load, assigns each plugin to an aggregation group (``necessary'' or ``voting''),
defines the timeout for plugins, etc.


\subsection{Plugin API}
\label{sec:design-plugins}
\name defines a robust plugin API that allows plugins to be queried with host data and a certificate chain and then return a response.
For authentication systems to be used with \name, they must be implemented as plugins and conform to the \name plugin API.
We provide both an asynchronous plugin API and a synchronous plugin API to facilitate the needs of different designs.

The synchronous plugin API is intended for use by simple authentication methodologies. Plugins using this API may optionally implement \texttt{initialize} and \texttt{finalize} functions for any setup and cleanup it needs to perform. For example, a plugin may want to store a cache or socket descriptor for long-term use during runtime. Each plugin must also implement a \texttt{query} function, which is passed a data object containing a query ID, hostname, IP address, port, certificate chain, and other relevant context. The certificate chain is provided to the plugin DER encoded and in openssl's \texttt{STACK\_OF(X509)} format for convenience. The \texttt{query} function returns the result of the plugin's validation of the query data (valid, invalid, abstain, or error) back to the policy engine.

The asynchronous plugin API allows for easier integration with more advanced designs, such as multithreaded and event-driven architectures. This API supplies a callback function through the \texttt{initialize} function that plugins must use to report validation decisions, using the query ID supplied by the data supplied to \texttt{query}. Thus the \texttt{initialize} function is required so that plugins may obtain the callback pointer (the \texttt{finalize} function is still optional). Asynchronous plugins also implement the \texttt{query} function, but return a status code from this function immediately and instead report their validation decision using the supplied callback.

\begin{figure*}
\centering
\includegraphics[width=0.7\textwidth]{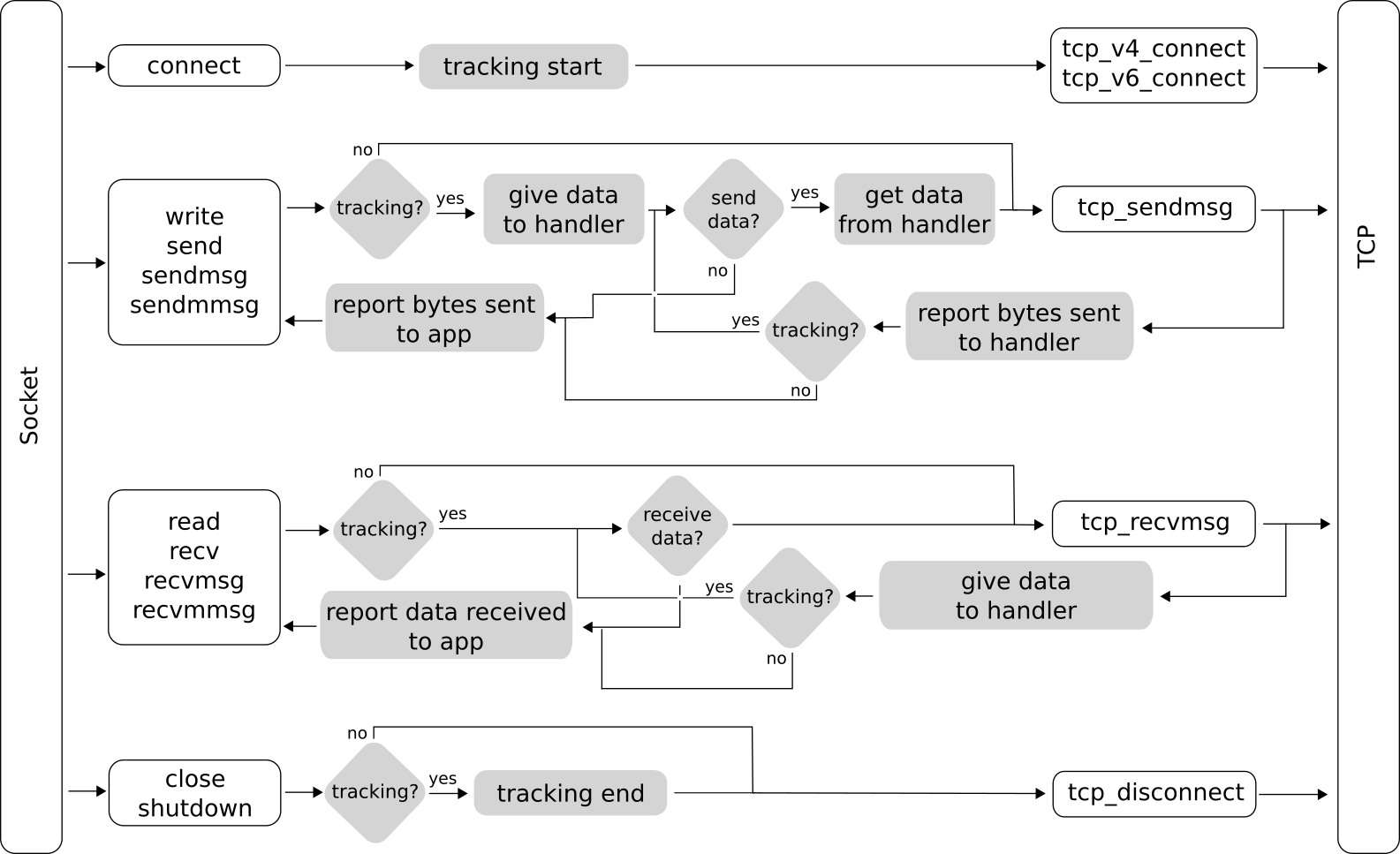}
\caption{Linux Traffic Interceptor simplified flowchart. Grey boxes correspond to hooks for handlers, white boxes are native system calls and kernel functions}
\label{fig:linux-interception}
\end{figure*}

\section{Linux Implementation}
\label{linux-implementation}

We have designed a research prototype implementation for \name in Linux. Source code is available upon request and will be provided publicly in the near future.

Our implementation uses a novel loadable kernel module (LKM) to intercept traffic at the
application layer, as data transits between the application and TCP.
No modification of native kernel code is required and the LKM can be loaded and unloaded at runtime.
Similarly to how Netfilter operates at the IP layer, \name
can intercept traffic at the socket layer, before data is delivered to TCP,
and pass it to application-level programs, where it can be (potentially) modified
and then passed back to the native kernel code for delivery to TCP. Likewise, interception can
occur when traffic arrives at TCP and before it is delivered to the application.
This enables \name to efficiently intercept TLS connections in the operating system
and validate certificates in the application layer.

The following discussion highlights the salient features of our implementation.


\subsection{Traffic Interceptor}
\label{sec:linux-traffic-interceptor}
\name provides generic traffic interception, by capturing traffic between sockets and the TCP protocol.
This is done by hooking several kernel functions, and wrapping them to add traffic interception as needed.
An overview of which functions are hooked and how they are modified is given in Figure~\ref{fig:linux-interception}.
Items in white boxes on the left side of the figure are system calls.
Items in white boxes on the right side of the figure are the wrapped kernel functions.
The additional logic added to the native flow of the kernel is shown by the arrows and gray boxes in Figure~\ref{fig:linux-interception}.

When the \name LKM is loaded, it hooks into the native TCP kernel functions whose pointers are stored in the global kernel structures \texttt{tcp\_prot} (for IPv4) and \texttt{tcpv6\_prot} (for IPv6).
When a user program invokes a system call to create a socket, the function pointers within the corresponding protocol structure are copied into the newly-created kernel socket structure, allowing different protocols (TCP, UDP, TCP over IPv6, etc.) to be invoked by the same common socket API.
The function pointers in the protocol structures correspond to basic socket operations such as sending and receiving data, and creating and closing connections.
Application calls to \texttt{read}, \texttt{write}, \texttt{sendmsg}, and other system calls on that socket then use those protocol functions to carry out their operations within the kernel.
Note within the kernel, all socket-reading system calls (\texttt{read}, \texttt{recv}, \texttt{recvmsg}, and \texttt{recvmmsg}) eventually call the \texttt{recvmsg} function provided by the protocol structure. 
The same is true for corresponding socket write system calls eventually calling the provided \texttt{sendmsg} function.
When the LKM is unloaded, the original TCP functionality is restored in a safe manner.


From top to bottom in Figure~\ref{fig:linux-interception}, the functionality of the traffic interceptor is as follows.
First, a call to \texttt{connect} informs the handler that a new connection has been created, and the handler can choose to intercept its traffic.

Second, when a call is made to send data on the socket, the interceptor
checks with the handler to determine if it is tracking this connection. If so, it forwards the traffic to the
traffic handler for analysis, and the handler chooses what data (potentially modified by the handler), if any, to relay to TCP.
After attempting to send data, the interceptor informs the handler how much of that data was successfully placed into the kernel's send buffer and provides notification of any errors that occurred.
At this point the interceptor allows the handler to send additional data, if desired.
The interceptor then queries the handler for the return values it wishes to report to the application (such as how many bytes were successfully sent or an error value).
This value is then returned to the application.

Third, a similar, reversed process is followed for the reception of data from the network. If the interceptor is tracking
the connection it can choose whether to receive data from TCP. Any data received is reported to the handler, which can choose whether to
report a different value to the application.
Note that handlers are allowed to report arbitrary values to applications for the amount of data sent or received, including false values, to allow greater flexibility in connection handling.
For example, to provide more time to obtain and parse an incomplete incoming external response, a handler may indicate that it wants a non-blocking application to believe that zero bytes of a prior query were sent through the network, even though the full query may have been sent. After the handler has completed its operation it can report to a subsequent receive call from the application that bytes were received, and fill the application's provided buffer with relevant data.

Finally, a call to \texttt{close} or \texttt{shutdown} informs the handler that the connection is closed. Note that the handler may also choose to abandon tracking of connections before this point.

Handlers for various network observation and modification can be constructed by implementing a small number of functions, which will be invoked by the traffic interceptor at runtime.
These functions roughly correspond to the grey boxes in Figure~\ref{fig:linux-interception}.
For example, handlers must implement functions to receive data from applications, send data to TCP, indicate whether to continue or cease tracking of a connection, etc.
The traffic interceptor calls these functions to provide the handler with data, receive data from the handler to be forwarded to applications or remote hosts, and other tasks.
Such an architecture allows developers to implement arbitrary protocol handlers as simple finite state machines, as demonstrated by the TLS handler and opportunistic TLS handlers described in the succeeding subsection.

Another option for implementing traffic interception would have been to use the Netfilter framework, but this is not an optimal approach.
\name relies on parsing traffic at the application layer, but Netfilter intercepts traffic at the IP layer. 
For \name to be implemented using Netfilter, \name would need to transform IP packets into application payloads. 
This could be done either by implementing significant portions of TCP, including out-of-order handling and associated buffers, or passing traffic through the network stack twice, once to parse the IP packets for \name and once for forwarding the traffic to the application.
Both of these options are problematic, creating development and performance overhead, respectively.


\subsection{TLS Handler}
\label{sec:linux-handshake-handler}

\begin{figure}
\centering
\includegraphics[width=0.9\columnwidth]{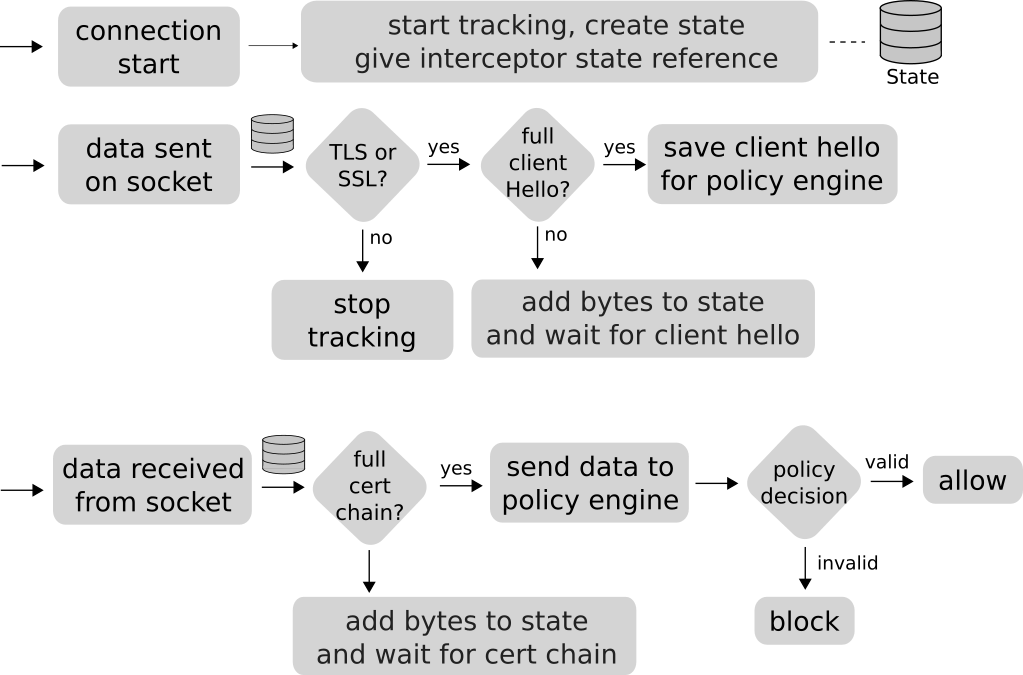}
\caption{Simplified view of TLS handler}
\label{fig:ssl-handler}
\end{figure}

\name includes a handler for the traffic interceptor dubbed the ``TLS handler''. The TLS handler extracts certificates from TLS network flows and forwards them to the policy engine for validation. Figure~\ref{fig:ssl-handler}
provides a high-level overview of how this handler operates. When a new socket is created, the handler creates state to track the connection, which the handler will have access to for all subsequent interactions with the interceptor. The destination IP address and port of the connection and PID of the application owning the connection are provided to the handler during connection establishment by the interceptor. Since the handler is implemented in an LKM, the PID of the socket can be used to obtain any further information about the application such as the command used to run it, its location, and even memory contents.
The TLS handler understands the TLS record and handshake protocols but does not perform interpretations of contained data. This minimizes additions to kernel-level code and allows ASN.1 and other parsing to be done in userspace by the policy engine.
When data is sent on the socket, the handler checks state data to determine whether the connection has initiated
a TLS handshake. If so, then it expects to receive a \texttt{ClientHello}; the handler saves this message for the policy engine so that it can obtain the hostname of the desired remote host, if the message contains a Server Name Indication (SNI) extension.
If SNI is not used, we use a log of applications' DNS lookups to infer the intended host,\footnote{In practice, all popular TLS implementations now use SNI, and so this fallback mechanism is almost never needed.} similar to work by Bates et al. \cite{bates2014securing}
When data is received on the socket, the TLS handler waits until it has received the full certificate chain, then it sends this chain and other data to the policy engine for parsing and validation.

\subsection{Opportunistic Encryption Handler}

We have also implemented an opportunistic encryption handler, which provides \name support for plaintext protocols that may choose to upgrade to TLS opportunistically, such as STARTTLS.
This handler performs passive monitoring of plaintext protocols (e.g. SMTP), allowing network data to be fast-tracked to and from the application and does not store or aggressively process any transiting data.
If at some point the application requests to initiate a TLS connection with the server (e.g. via a STARTTLS message), the handler processes acknowledgments from the server and then delivers control of the connection to the normal TLS handler, which is free to handle the connection as if it were conducting regular TLS.

It should be noted that the use of opportunistic TLS protocols by applications is subject to active attackers who perform stripping attacks to make the client believe the server does not support TLS upgrades, an existing vulnerability well documented by recent work \cite{durumeric2015neither,foster2015security,holz2016tls}. \name can prevent this type of attack, as discussed in Section~\ref{sec:assessment}.


\subsection{Policy Engine}
\label{sec:linux-policy-engine}

The policy engine receives raw certificate data from the TLS handler and then validates the certificates using the configured authentication services.
To avoid vulnerabilities that may arise from performing parsing and modification of certificates in the kernel, all such operations are carried out in user space by the policy engine.

Communication between \name kernel space and user space components is conducted via Netlink, a robust and efficient method of transferring data between kernel and user space, provided natively by the Linux kernel.
The policy engine asynchronously handles requests from the kernel module, freeing up the kernel threads to handle other connections while a response is constructed.




Native plugins must be written in either C or C++ and compiled as a shared object for use by the policy engine.
However, in addition to the plugin API \name supports an addon API that allows plugins to be written in additional languages.
Addons provide the code needed to interface between the native C of the policy engine and the target language it supports.
We have implemented an addon to support the Python language and have created several Python plugins.

\section{Security Analysis}
\label{sec:assessment}

The \name architecture, prototype implementation, and sample plugins have many implications for system security.
In this section we provide a security analysis of the centralized system design, application coverage, protection of applications from attackers, and protection of \name itself from attackers.

\subsection{Centralization}
Concentrating all certificate validation for a host into an operating system service has some risks and benefits. 
Any vulnerability in the service has the potential to impact all applications on the system. 
An exploit that grants an attacker root permission leads to compromise of the host. 
An exploit that causes a certificate to be rejected when it should be accepted is a type of denial-of-service attack. 
We first note that if an attacker is able to get \name to accept a certificate when it should not, any application that does its own certificate authentication correctly will be unaffected. 
If the application is broken, the \name failure will not make the situation any worse than it already was. The net effect is a lost opportunity to make it better.

The risks of centralization are common to any operating system service.
However, centralization also has a compelling upside.
For instance, all of our collective effort can be centered on making the design and implementation correct, and all applications can benefit.\footnote{All applications would likewise benefit from caching among authentication services.}
A single service is more scalable than requiring developers to secure each application or library independently.
It also enforces an administrator's preferences regardless of application behavior.
Additionally, when a protocol flaw is discovered, it can be more rapidly tested and patched compared to having to patch a large number of applications.

\subsection{Coverage}
Since one of the goals of \name is to enforce proper certificate validation on all applications on a system, the traffic interceptor is designed so that all TLS traffic on a system is visible to it.
Since the interceptor stands between the transport and application layers of the OS, it can intercept and access all TLS flows from local applications.
The handlers associated with the traffic interception component are made aware of a connection when a \texttt{connect} call is issued and can associate that connection with all data flowing through it.
Applications that utilize their own custom TCP/IP stack must utilize raw sockets, which require administrator privileges and are therefore implicitly trusted by \name.
Handlers can choose to read or ignore data at any time, modify or pass through data at any time, and continue or cease monitoring at any time.

To obtain complete coverage of TLS, our handlers need only monitor initial TLS handshakes (standard TLS) and the brief data preceding them (STARTTLS).
The characteristics of TLS renegotiation and session termination make this possible.
In TLS renegotiation, subsequent handshakes are encrypted with the parameters set by the preceding handshake, which depend on data from the certificate sent during that handshake. Thus if the policy engine correctly authenticates and validates the first handshake, TLS renegotiations are implicitly verified as well.
Attackers who obtained sufficient secrets to trigger a renegotiation, through some other attack on the TLS protocol or implementation (outside our threat model), have no need to take advantage of renegotiation as they have complete visibility and control over the connection already.
We also note that renegotiation is rare and typically used for client authentication for an already authenticated server, and has become less relevant for SGC or refreshing keys~\cite{ristic2014bulletproof}.

Session termination policies for TLS allow us to associate each TLS session with only one TCP connection.
In TLS, a close notify must be immediately succeeded by a responding close notification and a close down of the connection~\cite{rfc5246}.
Subsequent reconnects to the target host for additional TLS communication  are detected by the \name traffic interceptor and presented to the handlers.
We have found that TLS libraries and applications do indeed terminate a TCP session when ending a TLS session, although many of them fail to send an explicit TLS close notification and rely solely on TCP termination to notify the remote host of the session termination.

\subsection{Threat Analysis}
The coverage of \name enables it to enforce both proper and additional certificate validation procedures on TLS-using applications.
There are a variety of ways that attackers may try perform a TLS MITM against these applications. A selection of these methods and discussion of how \name can protect against them follows.
For each, we verified our solution utilizing an ``attacker'' machine acting as a MITM using sslsplit~\cite{sslsplit}, and a target ``victim'' machine running \name.
For some scenarios, the victim machine was implanted with our own CA in the distribution's shipped trust store, or the store of a local user or application.
Applications tested utilize the tools and libraries mentioned in section\ref{sec:benchmarking}.
\begin{itemize}
\item \textbf{Hacked or coerced Certificate Authorities:} Attackers who have received a valid certificate through coercion, deception, or compromise of CAs are able to subvert even proper CA validation. Under \name, administrators can choose to deploy pinning or notary plugins, which can detect the mismatch between the original and forged certificate, preventing the attacker from initiating a connection. We have developed plugins that perform these actions and verified that they prevent such attacks.
\item \textbf{Local malicious root:} Attackers utilizing certificates that have been installed into an application or user trusted store will be trusted by many target applications. Even Google Chrome will ignore certificate pins in the presence of a certificate that links back to a locally-installed root certificate. \name can protect against this by utilizing similar techniques to the preceding scenario.
\item \textbf{Absence of status checking:} Many applications still do not check OCSP or Certificate Revocation Lists to determine if a received certificate is valid~\cite{revocation}. In these cases attackers utilizing stolen certificates that have been reported can still perform a MITM.
Administrators who want to prevent this from happening can add an OCSP or CRL plugin to the policy engine and ensure these checks for all applications on the machine. We have developed an OCSP plugin and verified that it performs status checks where applicable.
\item \textbf{Failure to validate hostnames:} Some applications properly validate signatures from a certificate back to a trusted root but do not verify that the hostname matches the one contained in the leaf certificate. This allows attackers to utilize any valid certificate, including those for hosts they legitimately control, to intercept~\cite{georgiev2012most}.
The \name policy engine strictly validates the common name and all alternate names in a valid certificate against the intended hostname of the target host to eliminate this problem.
\item \textbf{Lack of validation:} For applications that blindly accept all certificates, attackers need only send a self-signed certificate they generate on their own, or any other for which they have the private key, to MITM a connection.
\name prohibits this by default, as the policy engine ensures the certificate  has a proper chain of signatures back to a trust anchor on the machine and performs the hostname validation described previously.
\item \textbf{STARTTLS downgrade attack:} Opportunistic TLS begins with a plaintext connection. A downgrade attack occurs when an active attacker suppresses STARTTLS-related messages, tricking the endpoints into thinking one or the other does not support STARTTLS. The net result is a continuation of the plaintext connection and possible sending of sensitive data (e.g., email) in the clear.
\name mitigates this attack by an option to enforce STARTTLS use.
When STARTTLS is used to communicate with a given service, \name records the host information.
Future connections to that host are then required to upgrade via STARTTLS.
If the host omits STARTTLS and prohibits its use, the connection is severed by \name to prevent leaking sensitive information to a potential attacker.\footnote{This could be further strengthened by checking DANE records to determine if the server supports STARTTLS. We are likewise interested in pursuing whether this technique can be used to protect against other types of downgrade attacks.}
\end{itemize}

\subsection{Hardening}

The following design principles strengthen the security of a TrustBase implementation.
First, the traffic interceptor and handler components run in kernel space. 
Their purpose is simple and limited so that overhead will be negligible. 
The handlers are implemented as finite state machines. 
Their small code size and limited functionality make it more likely that formal methods and source code auditing will provide greater assurance that an implementation is correct.
Second, the policy engine and plugins run in user space.
This is where error-prone tasks such as certificate parsing and validation occur.
The use of privilege separation~\cite{provos2003preventing} and sandboxing~\cite{maass2016systematic} techniques can limit the potential harm when any of these components is compromised. 
Third, plugins can only be installed and configured by an administrator, which prohibits unprivileged adversaries and malware from installing malicious authentication services.
Finally, communications between the handlers, policy engine, and plugins are authenticated to prevent local malware from spoofing a certificate validation result.


\name is designed to prevent a local, nonprivileged user from inadvertently or intentionally compromising the system. 
(1) Only privileged users can insert and remove the \name kernel module, prohibiting an attacker from simply removing the module to bypass it. 
The same is true for plugins.
(2) The communication between the kernel module component of \name and the user space policy engine is performed via a custom Generic Netlink protocol that protects against nonprivileged users sending messages to the kernel. 
The protocol defined takes advantage of the Generic Netlink flag GENL\_ADMIN\_PERM, which enforces that selected operations associated with the custom protocol can only be invoked by processes that have administrative privileges for networking (the capability mapped to CAP\_NET\_ADMIN in Linux systems). 
This prevents a local attacker from using a local netlink-utilizing process to masquerade as the policy engine to the kernel. 
(3) The policy engine runs as a nonroot, CAP\_NET\_ADMIN, chroot-jailed process that can be invoked only by a privileged user.
(4) The configuration files, plugin directories, and binaries for \name are write-protected 
to prevent unauthorized modifications from nonprivileged users. 
This protects against weakening of the configuration, disabling of plugins, shutting down or replacing the policy engine, or enabling of bogus plugins. 



\name aborts traffic interception for a given flow as soon as it is identified as a non-TLS connection.
Experimental results show that \name has negligible overhead with respect to memory and time while tracking connections. 
Thus it is unlikely that an attacker could perform a denial-of-service attack 
on the machine by creating multiple network connections, TLS or otherwise, any easier than in the non-\name scenario.
Such an attack is more closely associated with connection firewall policies.

An attacker may seek to compromise \name by crafting 
an artificial TLS handshake that results in some type of validation failure,  hoping to then subsequently launch a MITM attack against an application.
We attempt to shrink this possible attack surface by performing no parsing in the kernel except for TLS handshake records, 
which involves parsing nothing more than message type, length, and version headers.
ASN.1 and other data sent to the policy engine are evaluated and parsed by standard openssl functions, which have undergone widespread scrutiny and use for many years. 
Thus it seems a very difficult task to subvert \name by careful crafting of TLS message content.
Despite this we submit that the current implementation may not be foolproof in this respect and invite others to audit the code, which has been made public.
We note that while such an attack still requires more effort than if \name were not present, the presence of \name does not introduce additional vulnerabilities that do not already exist. For example, the target application must also improperly validate certificates for the attack to be successful.


\section{Evaluation}

We evaluated the prototype of \name to measure its performance, ensure compatibility with applications, and
test its utility for deploying authentication systems that can harden certificate validation.

\subsection{Performance}
\label{sec:benchmarking-2}

To measure the overhead incurred by \name, we instrumented our implementation to record the time required to establish a TCP connection, establish a TLS connection, and transferring a file of varying size (2MB - 500 GB).
We tested \name with two plugins, CA Validation and Certificate Pinning (see Section~\ref{sec:utility}).
The target host for these connections was a computer on the same local network as the client machine, to reduce the effect of latency and network noise.
In addition, the host presented the client with a valid certificate signed by an intermediate authority and finally a root authority (DigiCert). This ensured that the CA Validation plugin was forced to verify two certificates in the transmitted chain, which is a realistic circumstance for modern web browsing. The validity of the certificates also forced the plugins to execute all of their checks, rather than breaking out early due to detected problems.
Our testing was performed on a PC running Fedora 21 64-bit (kernel 3.17) with an i7-4790K CPU and 16 GB of RAM.
Times were recorded from the perspective of a local Python client application, using the \texttt{time} and \texttt{ssl} libraries.
Each measurement was performed 1,000 times.

\begin{figure}
\centering
\includegraphics[width=\columnwidth]{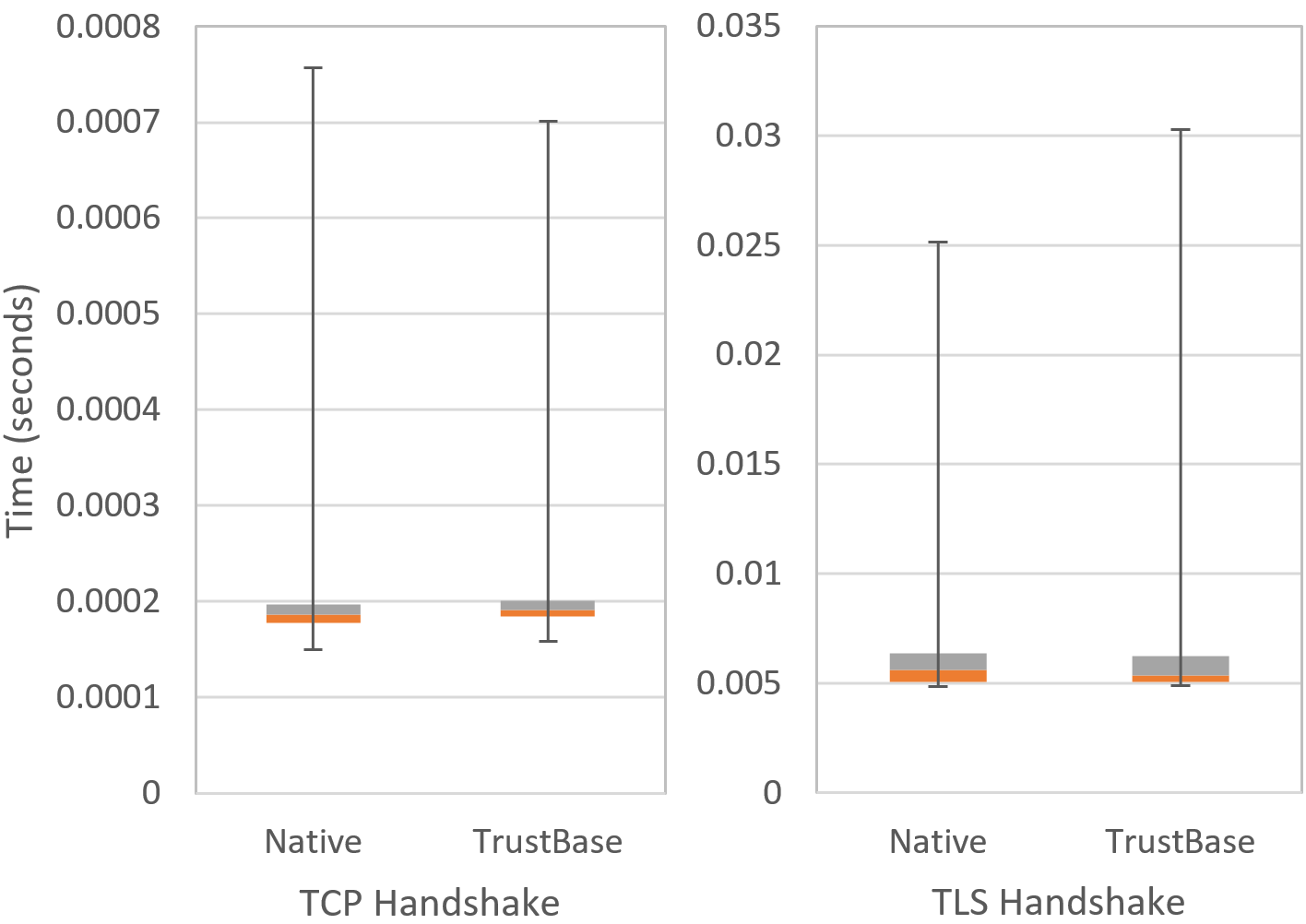}
\caption{Handshake Timings for TCP (left) and TLS (right) handshakes with and without \name running.}
\label{handshakes}
\end{figure}

Figure~\ref{handshakes} shows boxplots that characterize the timing of TCP and TLS handshakes, with and without \name active.
There is no discernible difference for TCP handshake timings and their average difference was less than 10 microseconds, with neither average consistently beating the other in subsequent experiments. This is expected behavior because the traffic interceptor is extremely light-weight for TCP connections.
Average TLS handshake times under \name and non-\name conditions also have an indiscernible difference, with average handshake times for this experiment of 5.9 ms and 6.0 ms, respectively. Successive experiments showed again that neither average consistently beat the other.
This means that the fluctuations of overhead incurred by kernel scheduling, network conditions, etc. account for more time than the additional control paths \name brings.
This is also expected, as the brevity of TLS handling code, its place in the kernel, the use of efficient Netlink transport and other design choices were made with performance in mind.
Since the latencies in these experiments were from two machines on the same local network, we can also expect that real-world connections would also have no discernible overhead when running \name.
The local network experiments allow to observe \name latency in a situation where its relative effect is maximized, rather than in situations where distance, hops, network conditions, and other factors account for the bulk of latency variations.

We also note that there was no additional difference when transferring files of various sizes.
This is intuitive since \name stops monitoring a connection once it has validated (or invalidated) its certificate, which happens before the first byte of data is sent.
Note that the \name timings for the TLS handshake may increase if a particular plugin is installed that requires more processing time or relies on Internet queries to function.

The memory footprint in our prototype is also negligible.
For each network connection, {\sname} temporarily stores only 212 bytes of data, plus the length of any TLS handshake messages encountered.
Connections not using TLS use even less memory than this and carry a zero-byte memory overhead once their nature has been determined and {\sname} ceases to monitor them.
This identification typically completes after the first byte of data is sent from the application.
All {\sname}-allocated memory in the kernel for a TLS connection is freed the moment the policy engine issues a decision and the traffic interceptor takes appropriate action.
Thus {\sname} on Linux requires very little memory per connection and all memory is usually freed within milliseconds of connection establishment, if not sooner.

\subsection{Compatibility}
\label{sec:benchmarking}

One goal of \name is to strengthen certificate authentication for existing, unmodified applications and to provide
additional authentication services that strengthen the CA system.
To meet this goal, \name must be able to enforce proper authentication behavior by applications, as defined by the system administrator's configuration. 
That is, applications are allowed to be more strict, but not less strict than \name in their validation decisions.
When the policy engine dictates that a certificate is valid, the application is allowed to accept it and proceed with a secure connection.
When the policy engine reports an invalid certificate, that connection should be aborted.

There are three possible cases for the policy engine to consider. (1) If a certificate has been deemed valid by both \name and the application, the policy engine allows the original certificate data to be forwarded on to the application, where it is accepted naturally.
(2) In the case where the application wishes to block a connection, regardless of the decision by \name, the policy engine allows this to occur, since the application may have a valid reason to do so.
We discuss in Section~\ref{sec:overriding}, the special case when a new authentication service is deployed that wishes to accept a certificate that the CA system normally would not.
(3) In the case where validation with \name fails, but the application would have allowed the connection to proceed, the policy engine blocks the connection by forwarding an intentionally invalid certificate to the application, which triggers any relevant validation errors an application may support, and then subsequently closes the connection.


We tested \name with 34 popular applications and libraries and tools, shown in 
Table~\ref{tab:validation-linux}.\footnote{These are a superset of the tools and libraries tested with CertShim}
\name successfully intercepted and validated certificates for all of them.
For each library tested, and where applicable, we created sample applications that performed no validation and improper validation (bad checking of signatures, hostnames, and validity dates).
We then verified that \name correctly forced these applications to reject false certificates despite those vulnerabilities.
In addition, we observed that \name caused no adverse behavior, such as timeouts, crashes, or unexpected errors.
\begin{table}
  \centering
  \footnotesize
\begin{tabularx}{\columnwidth}{@{}lXX@{}}
\toprule
\multicolumn{2}{l}{\textbf{Library}} & \textbf{Tool}\\
\midrule
\multicolumn{2}{l}{\small \textsc{C++}} & gnutls-cli \\
&libcurl & curl \\
&libgnutls & sslscan \\
&libssl & openssl s\_client \\
&libnss & openssl s\_time \\
\multicolumn{2}{l}{\small \textsc{Java}} & lynx \\
&SSLSocketFactory & fetchmail \\
\multicolumn{2}{l}{\small \textsc{Perl}} & firefox \\
&socket::ssl & chrome/chromium \\
\multicolumn{2}{l}{\small \textsc{PHP}} & mpop \\
&fsockopen & w3m \\
&php\_curl & ncat  \\
\multicolumn{2}{l}{\small \textsc{Python}} & wget \\
&httplib & steam \\
&httplib2 & thunderbird \\
&pycurl & kmail \\
&pyOpenSSL & pidgin \\
&python ssl \\
&urllib, urllib2, urllib3 \\
&requests \\
\bottomrule
\end{tabularx}
\vspace{10pt}
\caption{Common Linux libraries and tools compatible with {\sname}}
\label{tab:validation-linux}
\end{table}

\subsection{Android Compatibility}

To verify that the \name approach is also compatible with mobile applications,
we built a prototype for Android. Our Android implementation 
uses the \texttt{VPNService} so that it can be installed on an unaltered OS and
without root permissions. The drawback of this choice is that only one VPN service
can be active on the Android at a time.
In the long-term, adding socket-level interception to the Android kernel would be the right architectural choice,
and then \name could use similar traffic interception techniques as with the Linux implementation.

The primary engineering consequence of using the \texttt{VPNService} is that \name on Android intercepts IP packets from
applications but emits TCP (or UDP) packets to the network.  If it could use raw
sockets, then sname could merely transfer IP packets between the \texttt{VPNService} and the remote server.
Unfortunately, the lowest level socket endpoint an Android developer can create is the Java \texttt{Socket} or \texttt{DatagramSocket}, which encapsulate TCP and UDP payloads respectively. Therefore, we must emulate IP, UDP and TCP to facilitate communication between the \texttt{VPNService} and the sockets used to communicate with remote hosts.
For TCP, this involves maintaining connection state, emulating reliability, and setting appropriate flags (SYN, ACK, etc.) for TCP traffic.
Even with this requirement, overhead for \name on Android is also negligible.

To verify compatibility with mobile applications, we tested 16 of the most popular Android applications: Chrome, YouTube,
Pandora, GMail, Pinterest, Instagram, Facebook, Google Play Store,
Twitter, Snapchat, Amazon Shopping, Kik, Netflix, Google Photos, Opera,
and Dolphin. \name on Android successfully
intercepted and strengthened certificate validation for all of them.

\subsection{Utility}
\label{sec:utility}

To validate the utility of \name, we implemented six useful authentication services.
We discuss each of these and report statistics about their development. Times reported include the time developers spent to understand the service to be implemented.

\subsubsection{CA Validation}

The CA Validation service authenticates a certificate using native
\texttt{openssl} validation functions and standard practices for validating
hostnames, dates, etc.
This service enables \name to enforce proper CA validation.
In addition, this service allows \name to determine whether an alternative authentication service will make a different choice than the CA system,
which 
is useful when overriding the CA system (see the next section).
Implementing this service in C took approximately 12 hours and
contains 310 lines of code, including robust error handling.

\subsubsection{Whitelist}

The Whitelist service stores a set of certificates that are always
considered valid for their respective hosts.
This can be configured by
administrators to bypass authentication performed by other plugins, to speed up authentication for commonly-used certificates.
It also allows developers to deploy self-signed 
dummy certificates during testing that will bypass normal validation checks 
on the development machine instead of hard-coding an exception in the application.
Implementing this service in C took approximately 2 hours and
contains 128 lines of code.

\subsubsection{Certificate Pinning}

The Certificate Pinning service is similar to the Whitelist service, 
but uses Trust On First Use to pin certificates for any host.
The certificate received on subsequent visits to 
a host are checked against the certificate stored in a database to 
see if the certificates match. Expired certificates are 
replaced by the next certificate received by a 
connection to that domain.
This service could theoretically be extended to support a more robust pinning scheme such as TACK~\cite{tack}.
Implementing this service in C took approximately 5 
hours and contains 132 lines of code.

\subsubsection{Certificate Revocation}

The Certificate Revocation service checks the OCSP extension on a
certificate and, if present, calls the OCSP service to check whether
the certificate has been revoked. A recent paper by Liu et. al. shows that 8\% of certificates have
been revoked and that browsers often do not bother to check revocation
status \cite{revocation}. This service enables a user or organization to centralize and
enforce certificate revocation on all applications. To enforce
revocation, the \name configuration file simply needs to list the
revocation service in the {\em necessary} group.
Implementing this service in Python took approximately 8 hours and contains
99 lines of code.

\subsubsection{DANE}

The DANE service
uses the DNS system to distribute public keys in
a TLSA record \cite{hoffman2011using}. DANE enables
hosts to use self-signed certificates or to specify a particular set of valid
certificates, to strengthen the CA system against unauthorized authorities signing a certificate without the host's permission.
To implement this service, the plugin uses the certificate chain, port, and hostname, which are then used to perform a DNS query for the TLSA record.
Implementing this service in Python took approximately 8 hours
and contains 56 lines of code.

\subsubsection{Notary}

The Notary service is based on ideas presented by Perspectives~\cite{wendlandt2008perspectives} and Convergence~\cite{marlinspike2011ssl}.
It connects securely to one or more notary servers to validate the 
certificate received by the client is the same one that is seen by the notaries.
This can likewise be used to strengthen the CA system, or it can be used to validate self-signed certificates.
The service requires a configuration file that lists known notary servers and pinned certificates for each server.
The client hashes the leaf certificate for the connection in question and randomly chooses which notary server to act as a proxy to perform all queries to other known notary servers. The client then passes that notary the host, port, and hash of the leaf certificate for the connection. The notaries then respond with their validation verdicts, which are forwarded to the client, where they tallied.
Implementing the client side of this service in Python took approximately 7 hours and contains 60 lines of (client) code.

\section{Discussion}
\label{sec:discussion}

In this section we discuss a number of additional issues related to \name and its portability to additional operating systems,
compatibility with certificate pinning, and ability to override the CA system. We also provide a detailed comparison
of \name to CertShim.

\subsection{Operating System Support}

We designed the \name architecture so that it could be implemented
on additional operating systems. The main component that may need to
be customized for each operating system is the traffic interception module. We
are optimistic that this is possible because the TCP/IP networking
stack and sockets are used by most operating systems.

On Windows the traffic interception module could use the Windows
Filtering Platform API. This could be used to create a privileged,
kernel-mode driver that intercepts traffic between the TCP and
application layers.

Mac OSX provides a native interface for traffic interception between the TCP and socket levels of the operating system. Apple's Network Kernel Extensions suite provides a ``Socket Filter'' API that could be used as the traffic interceptor.

For iOS, Apple provides a Network Extension framework that includes a variety of APIs for different kinds of traffic interception. The App Proxy Provider API allows developers to create a custom transparent proxy that captures application network data. Also available is the Filter Data Provider API that allows examination of network data with built-in ``pass/block'' functionality.

Because Android uses a variant of the Linux kernel, we believe our
Linux implementation could be ported to Android with relative ease.

\subsection{Certificate Pinning}







Some applications have already begun to implement certificate pinning to provide greater security, rather than using the CA system
to validate a certificate. \name wants to avoid the situation where its authentication services declare a certificate
to be invalid when the application has validated it with pinning.
Our measurements indicate this is rare and affects relatively few applications, since the problem only arises when a
certificate does not validate by the CA system (e.g. a self-signed certificate shipped with the application).
In the short term, \name solves this problem by using the configuration file to whitelist programs that should skip using \name.
In the long term, this problem is solved by applications migrating to the \name for validation, as discussed above.

\subsection{Overriding the CA System}
\label{sec:overriding}

In some cases a system administrator may want to distrust the CA system entirely and rely solely on alternative authentication services.
For example, the administrator may want to allow CA-using applications to accept self-signed certificates that have been validated using a notary system such as Convergence, or she may want to use DANE with trust anchors that differ from
those shipped with the system. When this occurs, \name will use the new authentication
service and determine the certificate is valid, but applications using the CA system
will reject it, and validation will fail. We stress that this is not intended to override strong
certificate checks done by a browser (e.g. when talking to a bank), but to provide a path for
migrating away from the CA system when stronger alternatives emerge.

To handle these cases, \name provides two options: (1) migrating applications
to \name over the long term, and (2) a local TLS proxy that can coerce existing applications to
use new authentication services. We describe both of these options below.

\subsubsection{Migrating Applications}

The preferred option for overriding the CA system is to modify applications to rely on \name for
certificate validation, rather than performing their own checks.
This is facilitated by an API that applications can use to communicate with the \name policy
engine.
Applications can use the API to provide preferences and additional data to \name, such as requesting a self-signed certificate to be pinned, if allowed by the administrator.
The API also allows the application to receive validation error messages from \name, allowing it to display errors
directly in the application (\name displays notifications through the operating system).
Deferring to \name for certificate validation means that the administrator can configure any combination
of authentication systems she prefers.




\subsubsection{Proxying TLS Connections}
\label{sec:proxying}

An alternative option to override the CA system is to use a
local TLS proxy.
\name gives the administrator the option of running a proxy, but only in those cases where it
is absolutely needed, namely when the policy engine determines a certificate is valid but the CA system would reject it.
Note that in most cases this is not needed---for example even with Convergence running today, the
certificates validated by Convergence would likely also be validated by the CA system unless the
certificate was self-signed.
When the policy engine dictates that a connection should be proxied in this fashion, \name transparently redirects traffic to
a user space, nonprivileged TLS proxy daemon, and the daemon marshals traffic between the application and the intended remote host.
The application accepts the proxied connection because the policy engine dynamically generates a certificate that validates successfully
against a \name CA certificate, which can optionally be placed in the system trust store during installation.
The root certificate is uniquely generated (and periodically regenerated) locally and never leaves the machine. Its corresponding private key is also not readable by nonprivileged users.
Administrators should only use this feature if they are comfortable with the possibility of some of their secure traffic being proxied locally.

\subsection{Comparison to CertShim}

Because CertShim also tries to secure existing applications and introduce new authentication services, we provide a detailed description
of the differences here.
CertShim uses the \texttt{LD\_PRELOAD} environment variable in Linux systems to hook an application's use of some system calls and various TLS-related functions in common security libraries (e.g. openssl, gnutls, libnss).

CertShim has an advantage over \name in that it does not need to perform double validation for cases where an application is already performing certificate validation correctly. While \name can enforce proper and alternative validation techniques, any protected applications that do this already will do it a second time after \name has validated the secure connection. In addition, CertShim's wrapping of validation functions means that it can more easily override the CA system when administrators want to do this for existing applications, though this will only work with applications
that CertShim supports.

\name has some advantages that set it apart from CertShim in several notable ways:

  \subsubsection{Coverage}

  CertShim is not intended to protect browsers, and it cannot perform validation for applications that use a custom or unsupported security library, or have statically linked any security library. 
    In addition, it doesn't work with applications invoked by environment-passing exec functions that use custom environments, despite the use of \texttt{LD\_PRELOAD} in the parent environment (e.g., /etc/environment, .bashrc). In contrast, \name intercepts all secure traffic and thus can independently validate certificates for all applications, regardless of what library they used, how they were compiled, what user ran them, or how they were spawned.

\subsubsection{Maintenance}

    Since each security library has a distinct API, CertShim must be updated to support new libraries as they are released.
    Moreover, CertShim must use data structures internal to the security libraries it supports to obtain information such as hostname
    or port that is not always passed to the certificate validation functions.
    Libraries change their internals with surprising frequency; the current versions of PolarSSL (now mbed TLS) and GnuTLS are no longer compatible with CertShim, one year after its release. Maintaining backward compatibility with earlier library versions, and with different versions supported by different OS releases, makes updates an increasingly complex task.

    Another complication for CertShim is that some applications use security libraries in a hybrid fashion.
    For example, Chromium/Chrome uses a statically-linked version of its in-house BoringSSL for TLS connections and crypto operations, but libNSS is used for certificate validation. Similarly, Firefox has migrated away from libNSS for certificate validation to a statically-linked custom library (mozilla::pkix), but uses other security libraries for crypto. To maintain compatibility with hybrid approaches, CertShim would need to associate data from one library with data from another, and this may be impossible in cases involving static linking.  In contrast, to validate all certificates \name only needs to maintain compatibility with the TLS specification and the signatures of high-level functions of TCP in the Linux kernel. As a datapoint, the latter has had only two minor changes since Linux 2.2 (released 1999)---one change was to add a parameter, the other was to remove it.
    \name authentication services only need to call the public API of security libraries, making them immune to internal library changes.

\subsubsection{Administrator Control}

\name adopts the philosophy that system administrators should be in control of certificate validation on machines, mirroring their responsibilities for other security-related policies.
Using a loadable kernel module to intercept all secure traffic is in line with this philosophy---only administrators may load or unload its functionality, and, once loaded, every secure application is subject to the policies configured for \name. CertShim does not follow this philosophy; guest users and applications can easily opt out of its security policies. For example, guest users can remove CertShim from their \texttt{LD\_PRELOAD} environment variable,  and developers can bypass CertShim by statically-linking with security libraries, using an unsupported TLS library, or spawning child processes without CertShim in their environment.

\subsubsection{Local Adversary Protection}

CertShim's attack model does not assume a local adversary, wherein a nonprivileged, local, malicious application attempts to bypass validation. This can be done simply by modifying \texttt{LD\_PRELOAD}---malware commonly uses the same technique as CertShim \cite{ligh2014art}. \name mitigates such a scenario with its protected Netlink protocol, privileged policy engine, protected files, and kernel module that cannot be removed by a nonprivileged user. Recent studies of TLS MITM behavior suggest that local malware acting as a MITM is more prevalent than remote MITM attackers \cite{huang2014analyzing,tlsproxies}. This suggests that protection against such malware is an important part of certificate validation services.

\subsubsection{STARTTLS Enforcement}

Since CertShim hooks into TLS library calls, it cannot be invoked in cases where TLS is not present.
The unique placement of \name allows it to intercept connection data from both insecure and secure traffic, allowing it to record what hosts offer opportunistic STARTTLS services and recognize when a host suspiciously omits that offer. This allows \name to enforce STARTTLS usage, significantly reducing the attack surface for downgrade attacks.

\subsubsection{Additional Context for Plugins}

As alternative methods of certificate validation become more diverse, they may need to
access additional data from a TLS context, including data not normally associated with such connections. CertShim can only access information passed to standard network functions (such as \texttt{connect}, \texttt{gethostbyname}) and data from security libraries. However, the position of \name in the kernel allows it to have an immense breadth of contextual data that can be easily sent to the policy engine as additional information for authentication services. This includes but is not limited to a program's name, path, permissions, owning user, socket information, internal memory, etc. 



\section{Conclusion}

\name is able to secure existing applications, strengthen the CA system, and
provide simple deployment of new authentication services.
By using traffic interception, we have developed the first architecture that enables system administrators to ensure correct certificate validation for
all applications that use TLS, on a wide variety of operating systems.
Development and deployment of improved authentication services is aided by a 
policy engine, plugin API, and extensible support for additional programming languages.
We evaluated a research prototype for \name on Linux and demonstrated its negligible overhead and compatibility with
existing applications. To demonstrate the utility of \name, we have developed a variety of authentication services
with relatively little programming time required.  \name is a boon for both system administrators and researchers.
Administrators can rest assured that all secure connections are properly authenticated, and researchers can easily deploy and test authentication services they design.

\bibliographystyle{IEEEtranS}
\bibliography{IEEEabrv,bib}

\end{document}